\documentstyle[12pt]{article}

\newcommand{\EQ}{\begin{equation}}
\newcommand{\EN}{\end{equation}}
\newcommand{\EQA}{\begin{eqnarray}}
\newcommand{\ENA}{\end{eqnarray}}
\newcommand{\BA}{\begin{array}}
\newcommand{\EA}{\end{array}}
\newcommand{\NN}{\nonumber}
\newcommand{\spz}{\hspace{0.7cm}}

\begin{document}

\topmargin 0pt
\oddsidemargin 5mm
\newcommand{\NP}[1]{Nucl.\ Phys.\ {\bf #1}}
\newcommand{\PL}[1]{Phys.\ Lett.\ {\bf #1}}
\newcommand{\NC}[1]{Nuovo Cimento {\bf #1}}
\newcommand{\CMP}[1]{Comm.\ Math.\ Phys.\ {\bf #1}}
\newcommand{\PR}[1]{Phys.\ Rev.\ {\bf #1}}
\newcommand{\PRL}[1]{Phys.\ Rev.\ Lett.\ {\bf #1}}
\newcommand{\MPL}[1]{Mod.\ Phys.\ Lett.\ {\bf #1}}
\newcommand{\JETP}[1]{Sov.\ Phys.\ JETP {\bf #1}}
\newcommand{\TMP}[1]{Teor.\ Mat.\ Fiz.\ {\bf #1}}
\newcommand{\SL}{$SL(2,{\cal R})$}
\newcommand{\IN}{\int d^2x}
\newcommand{\bz}{{\bar z}}
\newcommand{\bpartial}{{\bar \partial}}
\newcommand{\bc}{{\bar c}}
\newcommand{\bb}{{\bar b}}
\newcommand{\bQ}{{\bar Q}}
\newcommand{\inv}{^{-1}}
\newcommand{\ta}{a^{\prime}_0}
\newcommand{\tab}{{\bar a}^\prime_0}
\newcommand{\tv}{v^\prime}
\newcommand{\tvb}{{\bar v}^\prime}
\newcommand{\SLC}{$SL(2,{\cal C})/SU(2)$}

\renewcommand{\thefootnote}{\fnsymbol{footnote}}

\newpage
\setcounter{page}{0}
\begin{titlepage}
\begin{flushright}
UCSBTH-91-55
\end{flushright}
\vspace{0.3cm}
\begin{center}
{\large
{\bf Extra Observables in Gauged WZW Models}} \\
\vspace{.7cm}
{\large  Nobuyuki Ishibashi
\footnote{Address after Nov. 1: Department of Physics, University of
California, Davis, CA 95616}} \\
\vspace{.5cm}
{\em
Department of Physics\\
University of California \\
Santa Barbara, CA 93106 } \\
\end{center}
\vspace{1.cm}

\begin{abstract}
It is known that Liouville theory can be represented as an \SL\ gauged
WZW model. We study a two dimensional field theory which can be obtained
by analytically continuing some of the variables in the \SL\ gauged WZW
model. We can derive Liouville theory from the analytically continued model,
( which is a gauged \SLC\ model, ) in a similar but more rigorous way than
from the original gauged WZW model. We investigate the observables of this
gauged \SLC\ model.  We find infinitely many extra observables which can not
be identified with operators in Liouville theory. We concentrate on
observables which are $(1,1)$ forms and the correlators of their integrals
over two dimensional spacetime. At a special value of the coupling constant
of our model,  the correlators of these integrals on the sphere coincide
 with the results from matrix models.

\end{abstract}
\vspace{.4cm}
\vspace{.4cm}
\centerline{October 1991}
\vspace{.3cm}
\end{titlepage}
\renewcommand{\thefootnote}{\arabic{footnote}}
\setcounter{footnote}{0}

\newpage
\section{Introduction}
\spz
Polyakov's discovery of the existence of $SL(2)$ current algebra
in two dimensional quantum gravity in the light-cone gauge\cite{pol}
 and
the subsequent success of the derivation of scaling exponents\cite{kpz},
suggested that
two dimensional quantum gravity could be rewritten in such a way that the
 $SL(2)$ current algebra is more apparent.
The authors of \cite{as}\cite{bo} showed that two dimensional quantum
gravity in the light cone gauge could be represented by
an \SL\ WZW model with the
constraint
\EQ
J^-=1.
\EN
The ``soldering'' procedure in \cite{kp} and
the study of \SL\ Chern Simons theory explored in
\cite{ver}, exposed the relation between two dimensional quantum gravity
and \SL\ gauge theory consisting
of zweibein and spin connection. These works imply
that the \SL\ structure is not a peculiarity of the light cone gauge but a
more
fundamental feature of two dimensional quantum gravity.

Although the light cone gauge reveals the \SL\ structure, the most
convenient gauge of two dimensional quantum gravity is conformal
gauge. In conformal gauge, two dimensional quantum gravity is
described by Liouville theory, which is simpler and more useful than the
complicated light cone gauge action.
In \cite{o'r}, it was shown that Liouville theory could also be represented
as a constrained \SL\ WZW model classically. This time  the $SL(2)$ currents
should be constrained as
\EQ
J^+={\bar J}^-=\sqrt{\mu}.
\EN
In \cite{tsu}\cite{dvv}, a gauged WZW model realizing the above
constraints was analyzed. Liouville theory was derived from this \SL\
gauged WZW model at the quantum level. Therefore, in this formulation,
we can see the \SL\ structure is also hidden in Liouville theory.

However the analysis of \cite{tsu}\cite{dvv} is somewhat formal because
they are dealing with WZW model of a noncompact group.
In this paper, we will propose another model which represents Liouville
theory. This new model, which is a gauged \SLC\ model, is
obtained by analytically continuing some of
the variables in the \SL\ gauged WZW model.
The analytic continuation does not spoil the left and right
$SL(2)$ current algebras in the WZW model.
Because of this analytic continuation, the arguments of
\cite{tsu}\cite{dvv} go through more rigorously in this
 model.
We will study the observables
and their correlation functions in this model.

The organization of this paper is as follows. In section 2, we first
review
the analysis of \cite{tsu}\cite{dvv}. In order to make their arguments
more rigorous, we analytically continue the \SL\
 gauged WZW model to the gauged
\SLC\ model. Liouville theory is derived from this gauged \SLC\ model in
a more rigorous way than in the derivation in \cite{tsu}\cite{dvv}.
In section 3, observables in the gauged \SLC\ model are discussed.
It is natural to expect that all the observables correspond to Liouville
theory operators. However, we find that
there exist infinitely many extra observables,
 which cannot be identified with operators in
Liouville theory.
We concentrate on the observables with conformal weight $(1,1)$ and
their integrals over two dimensional spacetime. The correlators of
such integrals are calculated on the sphere.
If one chooses the coupling constant of our model so that it corresponds
to Liouville theory induced by $c=-2$ conformal field theory,
 these correlators coincide with the correlators of the
observables at the first critical point of the one matrix model.
In section 4, we present a brief discussion of our
results. Appendix is devoted to definitions and some useful formulas
about \SLC\ model.

\section{Gauged WZW Model and Liouville Theory}
\spz
Let us consider the \SL\ gauged WZW model with gauge fields
$A_z^+$ and $A_\bz ^-$ following \cite{o'r},
\EQA
I
=
kS_{WZW}(g)
&+&
\frac{k}{\pi}\IN
\{A_\bz ^-(tr(t^+\partial gg^{-1})-\sqrt{\mu })
\nonumber \\
& &
\hspace{15mm}
+A_z^+(tr(t^-g^{-1}\bpartial g)-\sqrt{\mu })
\NN \\
& &
\hspace{15mm}
+A_z^+A_\bz ^-tr(t^+gt^-g\inv )\}.
\label{gWZW}
\ENA
Here $g\in \mbox{\SL }$ and $S_{WZW}(g)$ is the action of the
WZW model
\EQ
S_{WZW}(g)=-\frac{1}{2\pi }\IN tr\partial g\bpartial g\inv
           +\frac{i}{12\pi}\int_B d^3xtr(g\inv dg)^3,
\label{WZW}
\EN
and $t^\pm =t^1\pm t^2$ are the generators of \SL\ with positive and
negative roots respectively.
The action $I$ is invariant under the gauge transformations
$
\delta g
=
-\epsilon^-t^+g-g\epsilon^+t^-,\;
\delta A_z^+
=
\partial \epsilon^+,\;
\delta A_\bz ^-
=
\bpartial \epsilon^-$.

In \cite{tsu}\cite{dvv}, it was shown that Liouville theory can be
deduced from this gauged WZW model.
Let us review their derivation of Liouville theory.
 $g\in \mbox{\SL }$ can be parametrized
via the Gauss decomposition
\EQ
g=
\left(\BA {cc}
       1&v\\
       0&1
      \EA \right)
\left(\BA {cc}
       e^\phi &0\\
       0&e^{-\phi }
      \EA \right)
\left(\BA {cc}
       1&0\\
       {\bar v}&0
      \EA \right).
\label{gaus}
\EN
In these coordinates, the action becomes
\EQA
I
&=&
\frac{k}{\pi}\IN \{
\partial \phi \bpartial \phi +e^{-2\phi}\bpartial v\partial {\bar v}
\NN \\
& &
\hspace{15mm}
+A_z^+(e^{-2\phi }\bpartial v-\sqrt{\mu })
+A_\bz ^-(e^{-2\phi }\partial {\bar v}-\sqrt{\mu })
 \\
& &
\hspace{15mm}
+e^{-2\phi }A_z^+A_\bz^-\}.\NN
\label{fract}
\ENA
The authors in \cite{tsu}\cite{dvv} derived Liouville theory from this
field theory of $\phi$, $v$, ${\bar v}$ and $A$'s.
\footnote{
As
was stressed in \cite{o'r}, the Gauss decomposition is
possible for an element near the identity in the group
manifold. In this case, the \SL\ matrix $
g=
\left( \BA {cc}
        a&b\\
        c&d
       \EA
\right),
$
with $d=0$ cannot be represented by eq.(\ref{gaus}). Therefore, there is
a subtlety in representing the \SL\ WZW model as
the field theory
of $\phi $, $v$ and ${\bar v}$.}

Liouville theory appears if one integrates out $v$, ${\bar v}$
and $A$'s in the partition function
\EQ
Z=\int \frac{\mbox{[}d\phi dvd{\bar v}dA\mbox{]}}{Vol.}
e^{-I}.
\EN
Here $Vol.$ denotes the gauge volume of the gauge transformation
\EQA
\delta {\bar v}
=
-\epsilon^+,
& &
\delta A_z^+
=
\partial \epsilon^+,
\NN \\
\delta v
=
-\epsilon^-,
& &
\delta A_\bz ^-
=
\bpartial \epsilon^-.
\label{gauge}
\ENA
Integration over the $v$'s and $A$'s was done by
choosing the gauge $v={\bar v}=0$ or equivalently shifting the
integration variable $A$'s by $A_z^+\rightarrow A_z^++\partial {\bar v},
\;A_\bz^-\rightarrow A_\bz^-+\bpartial v$. After doing so, we are left with
the following expression for the partition function:
\EQ
Z=\int\frac{\mbox{[}d\phi dvd{\bar v}dA\mbox{]}}{Vol.}
\exp\{-\frac{k}{\pi}\IN (\partial \phi\bpartial\phi
+e^{-2\phi}A_z^+A_\bz^--\sqrt{\mu }A_z^+-\sqrt{\mu}A_\bz^-)\}.
\EN
The $v$, ${\bar v}$ integration merely corresponds to an overall
constant.
Since the functional integration measure for $\phi ,\;v,\;{\bar v}$ is
defined by the norm
\EQA
\| \delta g\| ^2
&=&
\IN tr(g\inv \delta g)^2
\NN \\
&=&
2\IN \{(\delta \phi )^2+e^{-2\phi}\delta v \delta {\bar v}\},
\label{hnorm}
\ENA
The $v,\;{\bar v}$ integration divided by the gauge volume $Vol.$
\footnote{
$Vol.$ is defined by the functional integration over the gauge parameter
$\epsilon $ with the norm
$\|\epsilon \|^2=\IN \epsilon^+\epsilon^-$.}
 gives us the factor arising from the determinant which we formally write
as $\prod_xe^{-2\phi}$. The
measure for $A_z^+$ and $A_\bz^-$ is defined by the norm
\EQ
\|\delta A\|^2
=
\IN \delta A_z^+\delta A_\bz^-.
\EN
The integration over $A_z^+,\;A_\bz^-$ contributes the
inverse of
 $\prod_xe^{-2\phi}$. Naively this cancels the
determinant coming from the $v$ integration. Thus we obtain,
\EQ
Z=\int \mbox{[}d\phi \mbox{]}\exp\{-\frac{k}{\pi}\IN
(\partial\phi\bpartial\phi -\mu e^{2\phi})\}.
\label{naiv}
\EN
The partition function therefore is the same as the partition function
of Liouville theory.
 However, notice that
we are comparing the determinants of the operator
$e^{-2\phi}$ acting by multiplication on $v$ with the same operator
 acting by multiplication on $A$.
This is a situation analogous to the one we encounter in
the ghost number anomaly in string
theory. In that case, one should compare the determinant of an
operator acting by multiplication on the ghost with that of the same operator
acting on the antighost. Since the spins of the ghost and the antighost are
different, there is a nontrivial difference between the two determinants.
Since the spins of $v$ and $A$ are different, we expect that there is
also a nontrivial difference between
the two
determinants in our case.
\footnote{
In \cite{tsu},
such a difference was intentionally neglected.}
Assuming that the difference of the two determinants changes eq.(\ref{naiv})
into the form,
\EQ
Z=\int \mbox{[}d\phi \mbox{]}\exp\{-\frac{k'}{\pi}\IN
(\partial\phi\bpartial\phi -Q\phi\partial\bpartial\sigma
-\mu^{\prime} e^{2\phi +\sigma} )\},
\label{modi}
\EN
where the spacetime metric is $ds^2=e^\sigma dzd\bz$, we can
determine the values of $k'$ and $Q$ \`{a} la DDK\cite{DDK} as
$k'=k-2,\;Q=\frac{k-1}{k-2}$. The Virasoro central charge of the
Liouville theory is $c=\frac{3k}{k-2}+6k-2$. This is exactly the
relation between the level of the \SL\ current algebra and the Virasoro
central charge in \cite{kpz}.

In this way, the authors of \cite{tsu}\cite{dvv} showed that the
partition function of the gauged WZW model coincides with that of
Liouville theory. Here, the following two remarks are in order.

1) In the Gauss
decomposition eq.(\ref{gaus}) of $g\in \mbox{\SL}$, $v$ and ${\bar v}$
are real. Therefore the system of bosons $v$ and ${\bar v}$ involves
a negative signature kinetic term and the norm eq.(\ref{hnorm}) is not
positive definite.
The origin of such a negative kinetic term and norm
 is the noncompactness of \SL\ .
Accordingly the norms of $A$'s and the gauge parameter
$\epsilon$ fail to be positive definite.
Therefore, strictly speaking,
the
functional integral over $v$'s and $A$'s  discussed above is
not well defined.
One way to make the functional
integral over $v$ and ${\bar v}$ well defined is to
continue $v$ and ${\bar v}$ so that the norm and the kinetic term become
positive definite.
If $k>0$ ( which we assume in the following ), this
 amounts to regarding $v$ and ${\bar v}$ ( and accordingly $A_z^+$ and
$A_\bz^-$ ) as complex
conjugate to each other.
As is shown in the appendix, the action
eq.(\ref{fract}) with $v$ and ${\bar v}$ complex conjugate to each
other, can be considered as a gauged version of
 SL(2,{\cal C})/SU(2) model \cite{gaw}.
Therefore, strictly speaking, we should
do the analytic continuation in order to make the above calculation rigorous.
The analytic
continuation seems to be legitimate, if what we are dealing with is the
field theory of $\phi$, $v$ and ${\bar v}$. However, considering that we
are dealing with an \SL\ gauged WZW model, this analytic continuation seems
subtle, because the negative kinetic term of $v$'s stems from the
noncompactness of \SL\, which is an essential feature of the group \SL\
{}.

2) In the usual \SL\ WZW model ( or
SL(2,{\cal C})/SU(2) model ), the variables $v$, ${\bar v}$ and $\phi$
are all scalars and $A$ is a vector field. However, because of the
presence of the terms proportional to $\sqrt{\mu }$ in eq.(\ref{fract}),
the action is not even rotationally invariant under such spin assignments.
In order to make the theory conformally invariant, we should
change the assignments.
We have to take the left and right conformal weights of $\phi$, $v$ and
${\bar v}$ so that the currents
$
e^{-2\phi}\partial {\bar v},\;
e^{-2\phi}\bpartial v
$
have the left and right conformal weights $(0,0)$. This can be achieved
by ``twisting'' the model. Namely, as was done in \cite{as}\cite{bo}\cite{o'r},
we add to the
stress tensor a derivative of the zeroth component of the chiral \SL\
current:
\EQ
T^{\prime }_{zz}=T_{zz}+\partial J_z^0,\;
T^{\prime }_{\bz\bz}=T_{\bz\bz}+\bpartial J_\bz^0.
\label{twisg}
\EN
This corresponds to shifting the conformal weights of $\phi$, $v$ and
${\bar v}$ to $(0,0)$, $(1,0)$ and $(0,1)$ respectively. The conformal
weights of the $A$'s are taken to be $(1,1)$.
Accordingly we
should modify the action as
\EQA
I
&=&
\frac{k}{\pi}\IN \{
\partial \phi \bpartial \phi -\phi\partial\bpartial\sigma
+e^{-2\phi-\sigma}\bpartial v\partial {\bar v}
\NN \\
& &
\hspace{15mm}
+A_z^+(e^{-2\phi -\sigma }\bpartial v-\sqrt{\mu })
+A_\bz ^-(e^{-2\phi -\sigma }\partial {\bar v}-\sqrt{\mu })
\NN \\
& &
\hspace{15mm}
+e^{-2\phi -\sigma}A_z^+A_\bz^-\}.
\label{twac}
\ENA
In \cite{tsu}\cite{dvv}, such a
twisting was not mentioned at the stage of considering the gauged WZW
action eq.(\ref{gWZW}). However,
we should start the discussion from this twisted action in order to
define the gauged WZW model to be rotationally invariant.
This modified action depends explicitly on the conformal factor
$\sigma $ of the metric. However it is invariant under the Weyl
transformation
$\sigma\rightarrow\sigma +\epsilon,
\;\phi\rightarrow\phi-\frac{1}{2}\epsilon $ as in the case of
a Feigin-Fuchs boson.

The arguments above show that the analysis
of \cite{tsu}\cite{dvv} described in the first part of this section is
somewhat formal. In order to make it
more rigorous,
we should start from the
action eq.(\ref{twac}), with $v$ and ${\bar v}$ complex conjugate to
each other, instead of eq.(\ref{fract}).
However the analytic continuation does not seem to be legitimate,
considering that we are dealing with the \SL\ WZW model. Therefore we would
rather propose this ( twisted ) gauged \SLC\ model eq.(\ref{twac}) as a
new model related to the \SL\ gauged WZW model. In this model, the
analysis of \cite{tsu}\cite{dvv} goes through more rigorously.
Here, instead, we will use an alternative method to
deduce Liouville theory starting from this gauged SL(2,{\cal C})/SU(2)
model eq.(\ref{twac}).
In our method, we
can derive eq.(\ref{modi}) directly without any assumption, and it
gives a more rigorous derivation of Liouville theory from the gauged
SL(2,{\cal C})/SU(2) model eq.(\ref{twac}).

Let us consider the partition function of the gauged SL(2,{\cal C})/SU(2)
model \footnote{For notational simplicity, we will discuss this model
on the sphere with the conformal metric $ds^2=e^\sigma dzd\bz$. }
\EQ
Z=\int \frac{\mbox{[}d\phi dvdA\mbox{]}}{Vol.}
e^{-I}.
\label{part}
\EN
Now $I$ is the action in eq.(\ref{twac}) and $v$ and ${\bar v}$ are
complex conjugate to each other
. $Vol.$ denotes the volume of the gauge transformation
eq.(\ref{gauge}), with $\epsilon^+$ and $\epsilon^-$ being complex
conjugate to each other.
 We will integrate out $v$, ${\bar v}$ and $A$ in
eq.(\ref{part}) and obtain Liouville theory.
After the twisting mentioned above, the functional
integration measures for these variables are defined by the norm
\EQA
\|\delta v\|^2
&=&
\IN e^{-2\phi}\delta v\delta{\bar v},
\NN \\
\|\delta A\|^2
&=&
\IN e^{-\sigma }\delta A_z^+\delta A_\bz^-.
\label{vAmet}
\ENA

In order to integrate out $v$ and $A$, one should somehow take care of
the gauge invariance. Essentially, what we will do here
is to fix the gauge as $A_z^+=A_\bz^-=0$.
The gauge fixed action then becomes
\EQA
I
&=&
\frac{k}{\pi}\IN \{
\partial \phi \bpartial \phi -\phi\partial\bpartial\sigma
+e^{-2\phi-\sigma}\bpartial v\partial {\bar v}\}
\NN \\
& &
+\frac{1}{\pi}\IN (b\bpartial c+{\bar b}\partial {\bar c}),
\label{loc}
\ENA
and the theory becomes a system consisting of the twisted
SL(2,{\cal C})/SU(2) model and ghosts. In this form, our model is
solvable using the current algebra technique.
However there is one thing one
has to notice with such a gauge choice. One cannot choose such a gauge
globally on a compact Riemann surface.
Indeed, by expanding the $A$'s in terms of the eigenfunctions of the
Laplacian
on the surface, one can show that $A$'s can be decomposed as
\EQA
A_z^+
&=&
\partial {\bar \Lambda} +a_0e^\sigma ,
\NN \\
A_\bz^-
&=&
\bpartial \Lambda+{\bar a}_0e^\sigma.
\label{Lapl}
\ENA
Here $\Lambda $ and ${\bar \Lambda}$ are $(1,0)$ and $(0,1)$ forms
respectively and $a_0$ and ${\bar a}_0$ are constants. The second terms
in eq.(\ref{Lapl}) cannot be gauged
away. Therefore the gauge eq.(\ref{loc}) is
possible only locally. Of course, such global obstructions do not
matter when one is canonically quantizing the system and calculating
commutation relations of operators. Therefore, quantities such as
anomalous dimensions of operators can be reliably computed using the
current algebra technique available in the gauge eq.(\ref{loc}).
In order to derive Liouville theory, we will construct an alternative
form of the action depending explicitly on the moduli $a_0$.

Let us change
variables from $A$ to $\Lambda $,
${\bar \Lambda }$, $a_0$ and ${\bar a}_0$ in the functional integration
eq.(\ref{part}).
The partition function becomes
\EQ
Z=\int \frac{\mbox{[}d\phi dvd\Lambda da_0\mbox{]}}{Vol.}
det'(\Delta )e^{-I}.
\label{part2}
\EN
The integration measure for $\Lambda$ and $a_0$ are defined by the
norm
\EQA
\|\delta \Lambda \| ^2
&=&
\IN \delta \Lambda \delta {\bar \Lambda},
\NN \\
\|\delta a_0\| ^2
&=&
\IN e^\sigma \delta a_0 \delta {\bar a}_0.
\ENA
$\Delta $ denotes the Laplacian $-\bpartial e^{-\sigma}\partial$ on
$(0,1)$ forms. $det'\Delta $ is the Jacobian for the change of variables
$A\rightarrow \Lambda ,\;{\bar \Lambda}$.
The action $I$ is written as
\EQA
I
&=&
\frac{k}{\pi}\IN \{
\partial \phi \bpartial \phi -\phi\partial\bpartial\sigma
+e^{-2\phi-\sigma}
\bpartial (v+\Lambda )\partial ({\bar v}+{\bar \Lambda})
\NN \\
& &
\hspace{15mm}
+a_0(e^{-2\phi }\bpartial (v+\Lambda )-\sqrt{\mu }e^\sigma )
+{\bar a}_0(e^{-2\phi }\partial ({\bar v}+{\bar \Lambda })-\sqrt{\mu }e^\sigma
)
\NN \\
& &
\hspace{15mm}
+e^{-2\phi +\sigma}a_0{\bar a}_0\}.
\ENA
Since the functional integration measure for $v$ defined by eq.(\ref{vAmet})
is invariant under the transformation $v\rightarrow v+\Lambda $,
${\bar v}\rightarrow {\bar v}+{\bar \Lambda}$,
we can factorize the $\Lambda $ integration in
eq.(\ref{part2}), which cancels the gauge volume $Vol.$.
Eventually, one obtains the
following expression of the partition function
\EQ
Z
=
\int \mbox{[}d\phi dv da_0\mbox{]}det'(\Delta )e^{-I_{f}},
\label{part3}
\EN
\EQA
I_f
&=&
\frac{k}{\pi}\IN \{
\partial \phi \bpartial \phi -\phi\partial\bpartial\sigma
+e^{-2\phi-\sigma}
\bpartial v\partial {\bar v}
\NN \\
& &
\hspace{15mm}
+a_0(e^{-2\phi }\bpartial v-\sqrt{\mu }e^\sigma )
+{\bar a}_0(e^{-2\phi }\partial {\bar v}-\sqrt{\mu }e^\sigma )
\NN \\
& &
\hspace{15mm}
+e^{-2\phi +\sigma}a_0{\bar a}_0\}.
\label{fiac}
\ENA
$det'(\Delta )$  can be expressed by a ghost $c$ ( $(1,0)$ form ) and
an
antighost $b$ ( $(0,0)$ form ) and their complex conjugates as usual.
The sum of $I_f$ and the ghost action
$I_{gh}=\frac{1}{\pi}\IN (b\bpartial c+{\bar b}\partial {\bar c})$
gives us a gauge fixed
action with explicit
moduli dependence. Locally it is possible to gauge away the moduli $a_0$
in $I_f$ to obtain eq.(\ref{loc}).
We can construct the BRST charges
\EQA
Q
&=&
\oint dzJ_{BRST}
=
\oint dzc
(J_z^+-k\sqrt{\mu})
\NN \\
{\bar Q}
&=&
\oint d\bz{\bar J}_{BRST}
=
\oint d\bz{\bar c}
(J_\bz^--k\sqrt{\mu}),
\label{BRST}
\ENA
where
\EQA
J_z^+
&=&
k(e^{-2\phi -\sigma}\partial {\bar v}+a_0e^{-2\phi}),
\NN \\
J_\bz^-
&=&
k(e^{-2\phi -\sigma}\bpartial v+{\bar a}_0e^{-2\phi}).
\ENA

The functional integration over $v$, $a_0$ and ghosts in
 eq.(\ref{part3}) will be done using the following
trick. Let us further decompose the moduli $a_0e^\sigma$ and
${\bar a}_0e^\sigma$ as
\EQA
a_0e^\sigma
&=&
\partial {\bar f}+f_0e^{2\phi +\sigma},
\NN \\
{\bar a}_0e^\sigma
&=&
\bpartial f +{\bar f}_0e^{2\phi +\sigma}.
\label{dec}
\ENA
This can be done by considering the nondegenerate bilinear form
\EQ
\|\omega\| ^2=\IN e^{-2\phi -\sigma}\omega {\bar \omega}
,
\EN
on $(1,1)$ forms and a Laplacian $-\bpartial \partial e^{-2\phi -\sigma}$,
which is self-adjoint with respect to this bilinear form.
Eqs.(\ref{dec}) amount to the orthogonal decomposition of the $(1,1)$ form
$e^\sigma$ into the zero mode and nonzero modes of this Laplacian and its
complex conjugate. Nonzero mode parts are written as derivatives of a
$(1,0)$ form $f$ and $(0,1)$ form ${\bar f}$. The coefficients of the
zero mode $e^{2\phi +\sigma}$ are
\EQ
f_0=a_0\frac{\IN e^\sigma}{\IN e^{2\phi +\sigma}},\;
{\bar f}_0={\bar a}_0\frac{\IN e^\sigma}{\IN e^{2\phi +\sigma}}.
\EN
Inserting this decomposition into eq.(\ref{fiac}), we obtain
\EQA
I_f
&=&
\frac{k}{\pi}\IN \{
\partial \phi \bpartial \phi -\phi\partial\bpartial\sigma
+e^{-2\phi-\sigma}
\bpartial v'\partial {\bar v}'-\mu e^{2\phi +\sigma}\}
\NN \\
& &
\hspace{15mm}
+\frac{k}{\pi}\frac{(\IN e^\sigma )^2}{\IN e^{2\phi +\sigma}}
\ta \tab ,
\label{fiac2}
\ENA
where
\EQA
\tv =v+f,
& &
\ta =a_0-\sqrt{\mu}\frac{\IN e^{2\phi +\sigma}}{\IN e^\sigma},
\NN \\
\tvb ={\bar v}+{\bar f},
& &
\tab =
{\bar a}_0-
\sqrt{\mu}\frac{\IN e^{2\phi +\sigma}}{\IN e^\sigma}.
\label{tilde}
\ENA

A good thing about this form of the action is that $\tab $ and
$\tv$ decouple
from each other. We can do the integration over $v$ and
$a_0$
separately. The $a_0$ integration is just a simple
Gaussian integration
\EQ
\int \mbox{[}da_0\mbox{]}
\exp \{
-\frac{k}{\pi}\frac{(\IN e^\sigma )^2}{\IN e^{2\phi +\sigma}}
\ta \tab
\}
=
\mbox{const.}\times \frac{\IN e^{2\phi +\sigma}}{\IN e^\sigma}.
\label{simG}
\EN
The integration over $v$ can be evaluated by the standard anomaly
calculation \cite{gaw}:
\EQA
& &
\int \mbox{[}dv\mbox{]}
\exp
\{-\frac{k}{\pi}\IN e^{-2\phi -\sigma}\partial \tvb
\bpartial \tv \}
\NN \\
& &
\hspace{15mm}
=
\frac{\IN e^\sigma}{\IN e^{2\phi +\sigma}}
det'(\Delta )^{-1}
\exp\{ \frac{2}{\pi}\IN \partial \phi \bpartial \phi
-\frac{1}{\pi}\IN \phi\partial\bpartial\sigma\} .
\label{vint}
\ENA
Putting all these together, we obtain
\EQA
Z
&=&
\int \mbox{[}d\phi dv da_0\mbox{]}det'(\Delta )e^{-I_{f}}
\NN \\
&=&
\mbox{const.}\int \mbox{[}d\phi \mbox{]}e^{-I_{Liou.}},
\NN \\
& &
\hspace{10mm}
I_{Liou.}= \frac{k-2}{\pi}\IN \partial\phi\bpartial\phi
        -\frac{k-1}{\pi}\IN \phi\partial\bpartial\sigma
        +\mu '\IN e^{2\phi +\sigma}.
\label{Liou}
\ENA
Eq.(\ref{Liou}) is the partition function of Liouville theory with the
cosmological constant $\mu '=-\frac{k\mu}{\pi}$. Thus, the partition
function of the \SLC\ model coincides with that of Liouville theory.

We would like to conclude this section by several comments.

For $k>0$,
$\mu$ should be negative for the functional integral to be well defined.
This implies that the action in eq.(\ref{gWZW}) has an imaginary part
proportional to $\sqrt{\mu}$. It does not cause a serious problem in our
analysis, because eventually the $\sqrt{\mu}$ term is relevant only in the
gaussian integration
 eq.(\ref{simG}).

The Liouville theory obtained in eq.(\ref{Liou}) is not always relevant
to two dimensional quantum gravity. In two dimensional quantum gravity
the cosmological constant $\mu '$ should be coupled to
the lowest dimensional operator in the
matter theory dressed by gravity. Therefore, eq.(\ref{Liou}) is relevant
to quantum gravity for only special values of $k$.

It is intriguing to observe that
in the above derivation, the existence of the moduli $a_0$ is essential
to generate the cosmological term $\IN e^{2\phi +\sigma}$. This moduli
also play an essential role in the next section.

\section{Extra Observables}
\spz
In this section, we would like to discuss the observables and their
correlation functions in the gauged \SLC\ model proposed in the previous
section. In the light of its relation to Liouville theory, it is natural
to expect that every observable in this model corresponds to
an operator
in Liouville theory. However we will find that there exist infinitely
many extra observables which do not correspond to Liouville
theory operators.

Let us consider the gauged \SLC\ model in the gauge fixed form
eq.(\ref{fiac}). The observables in this gauge are determined by the
usual BRST procedure using the BRST charges in eq.(\ref{BRST}).
The only singular
operator product expansions of $J_z^+$ and $J_\bz^-$ with $\phi$, $v$,
and ${\bar v}$ are given by
\EQA
J_z^+(z)v(w,{\bar w})
&\sim &
\frac{1}{z-w},
\NN \\
J_\bz^-(\bz ){\bar v}(w,{\bar w})
&\sim &
\frac{1}{{\bar z}-{\bar w}}.
\ENA
Therefore, operators made out of only $\phi$ are BRST closed.
It can be proved that correlation functions
of such operators $V_i$ ( $i=1,\cdots ,n $ ) in the gauged \SLC\ model
 reduce to correlation functions in Liouville
theory
\EQA
<V_1\cdots V_n>
&=&
\int\mbox{[}d\phi dvda_0\mbox{]}det'(\Delta )e^{-I_f}V_1\cdots V_n
\NN \\
&=&
\mbox{const.}\int\mbox{[}d\phi\mbox{]}e^{-I_{Liou.}}V_1\cdots V_n,
\label{cor}
\ENA
following the same procedure in the previous section.

The operators of the form $e^{-2l\phi }$ are important in the
application
of Liouville theory to two dimensional quantum gravity. In the gauged
\SLC\ model such operators are highest weight operators of the \SL\
current algebra:
\EQA
J_z^+(z)e^{-2l\phi }(w)
&\sim &
\mbox{regular},
\NN \\
J_z^0(z)e^{-2l\phi }(w)
&\sim &
\frac{l}{z-w}e^{-2l\phi }.
\ENA
The conformal weights of such highest weight operators are
$-\frac{l(l+1)}{k-2}-l$
\footnote{ To be precise, this suggests that $e^{-l(2\phi +\sigma )}$ is
a $(
-\frac{l(l+1)}{k-2}-l,\;
-\frac{l(l+1)}{k-2}-l)$ form.}
which of course coincide with the values
evaluated in Liouville theory. When $l=-1$,
the conformal weight is $(1,1)$, which is consistent with the fact that
this operator corresponds to the volume element in quantum gravity.

Therefore, every operator in Liouville theory can be represented as a
BRST invariant operator in the gauged \SLC\ model.
If such an operator is a null observable in the gauged \SLC\ model,
 it will decouple from the other operators in Liouville theory as
can be seen in eq.(\ref{cor}).
Hence,
if all the observables in the gauged
\SLC\ model are made out of $\phi$, we can have a complete correspondence
between the nontrivial operators in Liouville theory and the observables
in the
gauged \SLC\ model. However, there exist observables which consist
 not only of $\phi$ but also of other fields in the gauged \SLC\ model.

We can construct such extra observables starting from the following
observation. The ghost fields $b,\;c$ in the gauge fixed action are used
 to express the determinant $det'(\Delta )$ ( on the sphere, for
example, )
\EQ
det'(\Delta )=\int\mbox{[}dbdc\mbox{]}b{\bar b}(z_0)e^{-I_{gh}}.
\EN
Here, a pair of antighosts is inserted to soak up the zero mode in the
ghost path integral. The insertion point $z_0$ can be taken
arbitrarily.
Since the antighosts are $(0,0)$ forms, they have
one zero mode on a surface of any genus. Such a zero mode does not appear in
the action $I_{gh}$ or in the BRST charge eq.(\ref{BRST}).
 Therefore we eliminate it
from the theory by inserting $b{\bar b}$ as above. This situation is
analogous to the treatment of the $\xi$ zero mode of the superghost
bosonization in superstring theory\cite{fms}.
This leads
 an analogue of ``picture changing''
\EQ
O\longrightarrow \{ Q,bO\} ,
\label{gpic}
\EN
for the left moving sector along with the right moving one.
 By this operation, we can construct a new observable
$\{ Q,bO\}$ from an observable $O$, if $O$ contains no $c$.
The new observable $\{ Q,bO\}$ is in the
form of a BRST exact operator. However since it is an anticommutator of
BRST operator with an operator including the antighost zero mode, it does
not necessarily decouple from the other observables as in the case of
the picture changing in superstring theory.

Notice that this picture changing operation does not change the
conformal weight of the observable, because $b$ is a $(0,0)$ form field
and $Q$ commutes with the Virasoro operators. Therefore we obtain a new
observable with the same conformal weight by this operation.
In superstring theory, the picture changing operation
generates infinitely many equivalent expressions of
one
observable. However, as we
will see, in our case the picture changing operation generates an
infinite number of distinct observables.
By applying this picture changing operation to the observables like
$e^{-2l\phi}$, we can obtain infinitely many
observables which contain not only $\phi$ but also $v$'s and $a_0$'s. We
are not sure if the observables constructed in such a way exhaust the
observables of our model.

Let us consider the correlation functions of such observables on the
sphere:
\EQ
<V_1\cdots V_N>
=
\int \mbox{[}d\phi dvda_0dbdc\mbox{]}e^{-I_f-I_{gh}}b\bb (z_0)
V_1\cdots V_N
\EN
Such correlation functions can be calculated as follows.
As in the previous section, it is convenient to rewrite everything in
terms of $\ta$ and $\tv$ in eq.(\ref{tilde}). While $\ta$ and $\tv$
decouple from each other in the action,
\EQ
I_f
=
\frac{k}{\pi}\IN \{
\partial \phi \bpartial \phi -\phi\partial\bpartial\sigma
+e^{-2\phi-\sigma}
\bpartial \tv\partial \tvb-\mu e^{2\phi +\sigma}\}
+\frac{k}{\pi}\frac{(\IN e^\sigma )^2}{\IN e^{2\phi +\sigma}}
\ta \tab ,
\EN
there appears the nontrivial interaction term $\IN e^{2\phi +\sigma}$.
This term can be
taken care of by the method employed in \cite{gtw}\cite{gl}, namely, by
integrating over the $\phi$ zero mode first. Since the $\phi$ zero mode is
coupled to $\tv$ and $\ta$  in our model, we will proceed as follows.
Let us introduce a spacetime independent integration variable
$\phi_0$ and couple it to $I_f$ as
\EQ
I^\prime_f
=
\frac{k}{\pi}\IN \{
\partial \phi \bpartial \phi -(\phi +\phi_0)\partial\bpartial\sigma
+e^{-2\phi-\sigma}
\bpartial \tv\partial \tvb
-\mu e^{2\phi +2\phi_0+\sigma}\}
+\frac{k}{\pi}\frac{(\IN e^\sigma )^2}{\IN e^{2\phi +\sigma}}
\ta \tab -2\phi_0.
\EN
The transformation $\delta\phi =\epsilon,\;\delta v=\epsilon v,\;
\delta {\bar v}=\epsilon {\bar v},\;\delta \ta =\epsilon \ta ,\;
\delta \tab =\epsilon \tab ,\;\delta\phi_0=-\epsilon$ leaves
\EQ
\mbox{[}d\phi dvda_0d\phi_0\mbox{]}e^{-I^\prime_f}
\EN
invariant. Suppose there exists $\alpha_i$ such that
$e^{-2\alpha_i\phi_0}V_i$ is invariant under the above
transformation. Then the correlation function can be written as
\EQ
<V_1\cdots V_N>
=
\int \frac{\mbox{[}d\phi dvda_0d\phi_0dbdc\mbox{]}}{V}
e^{-I^\prime_f-I_{gh}}b\bb (z_0)
\prod_i e^{-2\alpha_i\phi_0}V_i,
\EN
where $V$ is the volume of the above continuous symmetry of the path
integral. This formula is easily proved, if one fixes the symmetry by
setting
$\phi_0=0$. Here we will fix the symmetry by the condition $\IN
e^\sigma \phi=0$, which kills the zero mode of $\phi$. $\phi_0$ plays
the role of the $\phi$ zero mode. Then, after
integrating over $\phi_0$, one obtains the
following expression for the correlation function
\EQA
<V_1\cdots V_N>
&=&
\mbox{const.}\times
(\mu ')^{k-1+\Sigma\alpha_i}
\Gamma (-(k-1+\Sigma\alpha_i))
\NN \\
& &
\times
\int \mbox{[}d\phi dvda_0dbdc\mbox{]}\delta (\int e^\sigma \phi )
e^{-I_0-I_{gh}}b\bb (z_0)
V_1\cdots V_N(\IN e^{2\phi +\sigma})^{k-1+\Sigma\alpha_i},
\NN \\
& &
\label{rem}
\ENA
where
\EQ
I_0
=
\frac{k}{\pi}\IN \{
\partial \phi \bpartial \phi -\phi\partial\bpartial\sigma
+e^{-2\phi-\sigma}
\bpartial \tv\partial \tvb\}
+\frac{k}{\pi}\frac{(\IN e^\sigma )^2}{\IN e^{2\phi +\sigma}}
\ta \tab .
\EN

The $\alpha_i$ are the analogs of the scaling dimensions in Liouville theory.
Since
\EQA
Q
&=&
k\oint c(e^{-2\phi -\sigma}\partial \tvb
+\frac{\IN e^\sigma}{\IN e^{2\phi +\sigma}}\ta )
\NN \\
{\bar Q}
& =&
k\oint \bc (e^{-2\phi -\sigma}\bpartial \tv
+\frac{\IN e^\sigma}{\IN e^{2\phi +\sigma}}\tab ),
\ENA
the picture changing operation
eq.(\ref{gpic}) changes the scaling dimension $\alpha$ by $-\frac{1}{2}$.
The scaling dimension is one of the physical quantum numbers in our
theory, on which the correlation functions crucially depend.
Therefore the picture changing operation produces an infinite number of
 distinct
observables.

If $k-1+\Sigma\alpha_i$ is not a positive integer, eq.(\ref{rem}) is not
well defined. One needs the analytic continuation as was done in
\cite{gl} to define it.
Since we are not sure if there exist any justifications for
that in our case, we will restrict ourselves to the case when
$k-1+\Sigma\alpha_i$ is a positive integer. If $k-1+\Sigma\alpha_i$ is a
positive integer, the factor $\Gamma (-(k-1+\Sigma\alpha_i))$ is
divergent. This divergence comes from the volume of $\phi_0$. We will
replace $
(\mu ')^{k-1+\Sigma\alpha_i}
\Gamma (-(k-1+\Sigma\alpha_i))$ by $
\frac{(-\mu ')^{k-1+\Sigma\alpha_i}}{(k-1+\Sigma\alpha_i)!}\log
\frac{1}{\mu '}$ and interpret $\log \frac{1}{\mu '}$ as the volume of
$\phi_0$ as was suggested in \cite{difr}.

In principle one can compute any correlation function of observables
by performing the functional integral in eq.(\ref{rem}), which
amounts to successive Gaussian integrals.
Here we will concentrate on the following extra observables.
Starting from
$O_0=e^{2\phi +\sigma}$, let us define $O_n$ inductively as
\EQA
O_{n+1}(w,{\bar w})
&=&
\frac{1}{k\pi}\{ {\bar Q},[ Q,b{\bar b}O_n(w,{\bar w})]\}
\NN \\
&=&
\frac{1}{k\pi}\oint_{\bar w} d\bz {\bar J}_{BRST}\oint_w dzJ_{BRST}
b\bb O_n(w,{\bar w}).
\label{pic}
\ENA
$O_n\propto
(J_z^+-k\sqrt{\mu})^n(J_\bz^--k\sqrt{\mu})^ne^{2\phi+\sigma}$
and $O_n$ does not contain $c$. Therefore eq.(\ref{pic})
is a well defined
picture changing operation. Since the conformal weight of
$O_0=e^{2\phi +\sigma}$ is $(1,1)$, all of the $O_n$ are $(1,1)$
operators.
In the rest of this section, we will show that it is possible to
compute
explicitly correlation functions of
 $\sigma_n=\IN O_n$,
\EQ
< \sigma_{n_1}\cdots \sigma_{n_N}>
=
\int \mbox{[}d\phi dvda_0dbdc\mbox{]}e^{-I_f-I_{gh}}b\bb (z_0)
\IN O_{n_1}\cdots \IN O_{n_N}.
\EN
One should drop the integrals of three of $O_n$'s in order to fix the
$SL(2,{\cal C})$ invariance in the above correlation function. We are
interested in these observables, because $\sigma_n$ are in a sense
``descendants'' of $\IN e^{2\phi +\sigma}$. The area of spacetime
$\IN e^{2\phi +\sigma}$ is one of the most important observables, in the
application of
 Liouville theory to two dimensional gravity. Also $\sigma_n$ can
be added to the action as a marginal perturbation. Hence if one knows
all the correlators of $\sigma_n$'s, one is able to solve such perturbed
field theories exactly.

Since $O_0=e^{-2\phi +\sigma}$ has scaling dimension $\alpha =-1$,
$O_n$ has $\alpha =n-1$. In order for $k-1+\Sigma\alpha_i$ to be
a positive integer, $k$ should be an integer. For later convenience,
we will restrict $k$ to be an integer satisfying $k\geq 4$.

Correlation functions of $O_n$'s have a remarkable property which
originates from their definition eq.(\ref{pic}).
 Namely they satisfy
\EQ
<O_{n}(x)O_{m}(y)\cdots >
=
<O_{n-1}(x)O_{m+1}(y)\cdots > \; (n>0).
\label{bose}
\EN
The proof is given using the same argument as one uses for the
demonstration of Bose sea equivalence in superstring theory\cite{fms}.
Writing
$O_n(x)=\frac{1}{k\pi}\oint {\bar J}_{BRST}\oint J_{BRST}b\bb
O_{n-1}(x)$, the left hand side of the above formula becomes
\EQ
\int \mbox{[}d\phi dvda_0dbdc\mbox{]}e^{-I_f-I_{gh}}b\bb (z_0)
\frac{1}{k\pi}
\oint_x {\bar J}_{BRST}\oint_x J_{BRST}b\bb O_{n-1}(x)O_m(y)
\cdots .
\EN
Since
the point $z_0$ at which the antighost is inserted is arbitrary,
we will take it to coincide with $y$. Then by using the BRST invariance
of the other observables, we move the integration contours of BRST
currents so that they surround only $y$. Thus we obtain the right hand
side of eq.(\ref{bose}).

Eq.(\ref{bose}) is useful in reducing
correlation functions of $\sigma_n$'s
 to a form in which they are easily calculated.
Eq.(\ref{rem}) and the interpretation of the divergent gamma function
suggest that the following equation between the correlation functions
 holds,
\EQ
<\IN O_{n_1}\cdots \IN O_{n_l}>
=
\frac{(-\mu ')^{k-1+\Sigma -l}}{(k-1+\Sigma -l)!}
<\IN O_{n_1}\cdots \IN O_{n_l}(\IN O_0)^{k-1+\Sigma -l}>,
\label{hold}
\EN
where $\Sigma =\sum n_i$.
Eq.(\ref{bose}) implies
\EQ
<\IN O_{n_1}\cdots \IN O_{n_l}
(\IN O_0)^{k-1+\Sigma -l}>
=
<(\IN O_1)^{\Sigma}
(\IN O_0)^{k-1}>.
\label{01}
\EN
Therefore the computation of correlation functions of $\sigma_n$'s
is reduced to that
 of the correlation functions of $\sigma_1$'s and $\sigma_0$'s only.

Now we are going to evaluate the right hand side of eq.(\ref{01}). We
should fix the $SL(2,{\cal C})$ invariance to define this correlation
function.
Suppose we fix the positions of three of the $O_0$'s. This is
possible, since $k\geq 4$.
$O_1$ has the form
\EQ
\frac{1}{k\pi}\oint d\bz {\bar J}_{BRST}\oint dz J_{BRST}e^{2\phi +\sigma}
=
\frac{1}{\pi}\{ {\bar Q},\bb (\partial \tvb +
\frac{\IN e^\sigma}{\IN e^{2\phi +\sigma}}\ta e^{2\phi +\sigma})\}.
\EN
Hence $\IN O_1$ is written as
\EQ
\IN O_1
=
\frac{1}{\pi}\{ {\bar Q},\IN \bb \partial \tvb\}
+\frac{k}{\pi}\frac{(\IN e^\sigma )^2}{\IN e^{2\phi +\sigma }}\ta \tab .
\EN
The first term is a commutator of the BRST operator and an operator which
does not contain the antighost zero mode. It decouples from the other
observables in the correlation function. Hence the right hand side of
eq.(\ref{01}) is equal to
\EQ
<(
\frac{k}{\pi}\frac{(\IN e^\sigma )^2}{\IN e^{2\phi +\sigma }}\ta \tab )^\Sigma
(\IN e^{2\phi +\sigma})^{k-1}>.
\label{rhs}
\EN
It is annoying to have $\IN e^{2\phi +\sigma}$ in the denominator, but
they all disappear after the $a_0$ integration.
The integration over $a_0$ is easily done.
We find
\EQ
<(
\frac{k}{\pi}\frac{(\IN e^\sigma )^2}{\IN e^{2\phi +\sigma }}\ta \tab )^\Sigma
(\IN e^{2\phi +\sigma})^{k-1}>
=
\Sigma !<(\IN e^{2\phi +\sigma})^{k-1}>.
\EN
Using eq.(\ref{hold}), one finally obtains
\EQ
<\sigma_{n_1}\cdots \sigma_{n_l}>
=
(-\mu ')^{\Sigma -l}\frac{\Sigma !(k-1)!}{(k-1+\Sigma -l)!}Z.
\EN
Since the $\sigma_n$'s can be added to the original action as a
perturbation, this result makes it possible to calculate various
quantities exactly in such a perturbed theory.

The $k=4$ case is extremely
interesting. Since $Z\propto (-\mu ')^3\log \frac{1}{\mu '}$,
\EQ
<\sigma_{n_1}\cdots \sigma_{n_l}>
\sim
\frac{(-\mu ')^{3+\Sigma -l}\Sigma !}{(3+\Sigma -l)!}.
\log \frac{1}{\mu '}
\EN
Up to a constant and the Liouville volume $\log \frac{1}{\mu '}$,
these correlation functions precisely coincide with the correlation
functions of the first critical point of one matrix model on the sphere
\cite{mat}.

This coincidence is very suggestive.
$k=4$ is exactly the
point that our \SLC\ model realizes the Liouville theory which is
induced by the matter theory with $c=-2$. The first critical point of
the one matrix model is supposed to correspond to the quantum gravity
coupled to $c=-2$ matter theory.
$k=4$ is the only integer greater than three, for which
the corresponding Liouville theory is relevant to
two dimensional quantum gravity. For other values of $k$, $\mu '$ does
not couple to $\IN e^{2\phi +\sigma}$ in two dimensional quantum
gravity.

Unfortunately it seems that such coincidence does not exist for the
correlators of $\IN O_n$ on higher genus surfaces. In general, matrix
model correlators are not compatible with eq.(\ref{bose}).

\section{Conclusions and Discussions}
\spz
In this paper, we have studied the relationship between the gauged \SLC\
model and Liouville theory. One of the crucial points in our analysis is
that there exists moduli $a_0$ on
any compact Riemann surfaces. This is because the gauge field in our
model is a $(1,1)$ form.
The existence of the moduli $a_0$ is
essential to obtain the cosmological term in Liouville theory. Then we
 investigated the observables in the gauged \SLC\ model.
Although the partition function of this model coincides with that of
Liouville theory,
 the gauged \SLC\ model possesses more observables than
Liouville theory. The existence of the zero modes of antighosts ( which
is of course deeply related to the existence of $a_0$, ) was
important in the construction of such observables.
We have calculated the correlators of some of these
extra observables.
They look quite similar to correlators in the matrix
models and coincide with them when $k=4$.

We are not sure if our results have any implications for two dimensional
quantum gravity. The extra observables studied in section 3 do not exist
in Liouville theory. And their correlators do not appear to coincide
with the matrix model results on higher genus surfaces. Still it is
possible to conceive that some modified version of the gauged \SLC\ model
would reproduce the matrix model results completely and elucidate the
importance of the \SL\ structure in two dimensional quantum gravity.

It is straightforward to extend our analysis to the case of more general
constrained WZW models. For example, $SL(N,{\cal R})$ WZW model with
constraints similar to \SL\ case was shown to be relevant to Toda field
theories\cite{o'r}. Gauged WZW models based on certain super Lie groups
 yield super
Toda field theories\cite{ina}. In both of these cases, there exist $(1,1)$
gauge fields, and $(0,0)$ antighosts when one fixes the gauge.
It is possible to consider the construction of observables by the
``picture
changing operation'' using these antighost fields.
The generalizations to
these cases will be reported elsewhere.

\section{Appendix}
\spz
In this appendix we will give the definition and some useful properties
and formulas of the \SLC\ model and its gauged version.

The \SLC\
model\cite{gaw} is defined by the action
\EQ
I=
kS_{WZW}(gg^\dagger ),
\label{c/u}
\EN
where $g\in SL(2,{\cal C})$ and $S_{WZW}$ is the WZW action
eq.(\ref{WZW}).
This model describes the induced gauge theory which is obtained by
integrating the matter part
in G/H gauged
WZW model \cite{gaw}, when H is $SU(2)$.
The gauge transformation of such an induced gauge theory corresponds to
\EQ
g\longrightarrow gh,\;h\in SU(2).
\EN
In order to define the functional integral, we should fix this
invariance. This can
be done most conveniently by taking $g$ to be
\EQ
g=
\left( \BA {cc}
        1&v\\
        0&1
       \EA
\right)
\left( \BA {cc}
        e^{\frac{\phi }{2}}&0\\
        0&e^{-\frac{\phi }{2}}
       \EA
\right)
,
\EN
and
\EQ
gg^\dagger
=
\left(\BA {cc}
       1&v\\
       0&1
      \EA \right)
\left(\BA {cc}
       e^\phi &0\\
       0&e^{-\phi }
      \EA \right)
\left(\BA {cc}
       1&0\\
       {\bar v}&0
      \EA \right).
\label{gaus'}
\EN
Here $\phi$ is real, $v$ is complex, and ${\bar v}$ is the complex
conjugate of $v$. Eq.(\ref{gaus'}) is similar to the Gauss
decomposition eq.(\ref{gaus}).
Contrary to the Gauss decomposition eq.(\ref{gaus}), however,
$gg^\dagger$ for any $g\in SL(2,{\cal C})$ can be represented as
in eq.(\ref{gaus'}).
Inserting
this parametrization of $g$ into eq.(\ref{c/u}), one obtains
the action,
\EQA
I
&=&
\frac{k}{\pi}\IN \{
\partial \phi \bpartial \phi +e^{-2\phi}\bpartial v\partial {\bar v}
\},
\label{slcac}
\ENA
with $v$ and ${\bar v}$ complex conjugate to each other.
The \SLC\ model is exactly described by
this action.

This theory possesses $SL(2)$ chiral currents:
\EQA
& &
J_z^+=\frac{k}{\sqrt{2}}e^{-2\phi}\partial {\bar v}
\NN \\
& &
J_z^0=k(\partial \phi +e^{-2\phi}v\partial {\bar v})
\NN \\
& &
J_z^-=\frac{k}{\sqrt{2}}
(\partial v-2\partial \phi v-e^{-2\phi}v^2\partial {\bar v}),
\NN
\ENA
and
\EQA
& &
J_\bz^-=\frac{k}{\sqrt{2}}e^{-2\phi}\bpartial v
\NN \\
& &
J_\bz^0=k(\bpartial \phi +e^{-2\phi}{\bar v}\bpartial v)
\NN \\
& &
J_\bz^+=\frac{k}{\sqrt{2}}
(\bpartial {\bar v}-2\bpartial \phi {\bar v}
-e^{-2\phi}{\bar v}^2\bpartial v).
\label{rcu}
\ENA
These currents correspond
 to the transformation
\EQ
g\longrightarrow hg,\;h\in SL(2,{\cal C}),
\label{SLC}
\EN
and satisfy the left and right $SL(2)$ current algebras. They are
important in constructing the BRST charge in $G/SU(2)$ gauged WZW models.
The form of the currents in terms of $\phi$, $v$ and ${\bar v}$
are the same as the chiral $SL(2)$ currents in the \SL\ WZW model.
However, since $v$ and ${\bar v}$ are complex conjugate to each other,
the left and right currents are related via complex conjugation in a
different way in this model. The stress tensor can be written in terms
of these $SL(2)$ currents in the Sugawara form.

We will define the functional integral measure for $\phi$ and $v$ by the
norm
\EQ
\IN e^\sigma \{(\delta \phi )^2+e^{-2\phi}\delta v \delta {\bar v}\},
\label{mec/u}
\EN
which is invariant under eq.(\ref{SLC}).
If one
 performs the $v$ integration first in the $\phi$ background and then do
the $\phi$ integration,
the functional integral
\footnote{Since this system has a
global symmetry $g\rightarrow hg$, one should
divide by the volume of such a global symmetry to define this functional
integral.}
\EQ
\int \mbox{[}d\phi dv\mbox{]}e^{-I},
\label{suc}
\EN
is evaluated essentially by successive Gaussian integrals.

The $v$ integration in the $\phi$ background yields the following
partition function \cite{gaw}
\EQ
\int \mbox{[}dv\mbox{]}\exp
\{-\frac{k}{\pi}\IN e^{-2\phi}\partial {\bar v} \bpartial v\}
=
(\frac{det'(\Delta )}{\IN e^\sigma })^{-1},
\exp\{\frac{2}{\pi}\IN\partial\phi\bpartial\phi
+\frac{1}{\pi}\IN\phi\partial\bpartial\sigma\}.
\EN
where the measure $\mbox{[}dv\mbox{]}$ is defined by the norm
$\|\delta v\|^2=\IN e^{-2\phi +\sigma}\delta v\delta {\bar v}$.
Correlation functions of $v$'s are calculable using Wick's theorem.
After the $v$ integration,
the functional integral eq.(\ref{suc}) becomes a theory of boson $\phi$
with Feigin-Fuchs type action.
Hence, in
principle, one can calculate any correlation function in this model.
Indeed, computation of some of the correlation functions in this model
in this way was done in \cite{gaw}, and the results were consistent
with the Knizhnik-Zamolodchikov equations derived from the $SL(2)$ current
algebras satisfied by eq.(\ref{rcu}).

We can gauge \SLC\ model without any problem in a similar way to the
\SL\ case in eq.(\ref{gWZW}). We obtain exactly the action eq.(\ref{fract})
with $v$ and ${\bar v}$ complex conjugate to each other. Also one can
twist the \SLC\ model as was done in section two.
In this case, the action should be
modified as follows
\EQ
I
=
\frac{k}{\pi}\IN \{
\partial \phi \bpartial \phi -\phi\partial\bpartial\sigma
+e^{-2\phi -\sigma}\bpartial v\partial {\bar v}
\},
\EN
where now $v$ ( ${\bar v}$ ) is a $(1,0)$ ( $(0,1)$ ) form.
The $SL(2)$ currents in eq.(\ref{rcu}) with a
little modification
( changing all $\phi$'s in the expression to
$\phi +\frac{1}{2}\sigma$ )
still satisfy the $SL(2)$ Kac Moody algebra, after the twisting. The
Virasoro generators are written in terms of these currents as
in eq.(\ref{twisg}). The twisted \SLC\ model is also solvable by successive
functional Gaussian integration in the same way as in the untwisted case.

\section*{Acknowledgements}
It is a pleasure to thank J. Horne, M. Li, S. Nojiri and A. Steif for
useful conversations and comments. This work is supported by the Nishina
Memorial Foundation and NSF Grant No. PHY86-14185.


\end{document}